\documentclass[reprint,superscriptaddress,showframe,nofootinbib,aps,prd,longbibliography,noeprint]{revtex4-2}

\newcommand{\mat}[1]{\textsf{\textbf{#1}}}

\newcommand{\arcsec}{{^{\prime\prime}}}

\newcommand{\bigo}[1]{\mathcal{O}(#1)}
\newcommand{\mrm}[1]{\mathrm{#1}}
\newcommand{\CL}{\textsc{CosmoLike}\ }
\newcommand{\dd}{\mathrm{d}}

\usepackage{graphicx}
\usepackage{dcolumn}
\usepackage{bm}
\usepackage{hyperref}
\usepackage{orcidlink}
\usepackage{amsmath}	
\usepackage{amssymb}
\usepackage{gensymb}
\usepackage{color}
\usepackage{multirow}
\usepackage{siunitx}
\definecolor{orange}{rgb}{1.0, 0.49, 0}

\begin{document}

\preprint{}

\title{Kinematic Lensing with the Dark Energy Spectroscopic Instrument---Probing structure formation at very low redshift}

\author{Jiachuan Xu\orcidlink{0000-0003-0871-8941}}
\email{jiachuanxu@arizona.edu}
\affiliation{Department of Astronomy/Steward Observatory, University of Arizona, 933 North Cherry Avenue, Tucson, AZ 85721, USA}

\author{Tim Eifler}
\affiliation{Department of Astronomy/Steward Observatory, University of Arizona, 933 North Cherry Avenue, Tucson, AZ 85721, USA}
\affiliation{Department of Physics, University of Arizona, 1118 E Fourth Street, Tucson, AZ 85721, USA}

\author{Eason Wang}
\affiliation{Department of Astronomy/Steward Observatory, University of Arizona, 933 North Cherry Avenue, Tucson, AZ 85721, USA}
\affiliation{Department of Physics, University of Arizona, 1118 E Fourth Street, Tucson, AZ 85721, USA}

\author{Elisabeth Krause}
\affiliation{Department of Astronomy/Steward Observatory, University of Arizona, 933 North Cherry Avenue, Tucson, AZ 85721, USA}
\affiliation{Department of Physics, University of Arizona, 1118 E Fourth Street, Tucson, AZ 85721, USA}

\author{Spencer Everett}
\affiliation{Jet Propulsion Laboratory, California Institute of Technology, Pasadena, CA 91109, USA}

\author{Eric Huff}
\affiliation{Jet Propulsion Laboratory, California Institute of Technology, Pasadena, CA 91109, USA}

\author{Pranjal R. S.}
\affiliation{Department of Astronomy/Steward Observatory, University of Arizona, 933 North Cherry Avenue, Tucson, AZ 85721, USA}

\author{Yu-Hsiu Huang}
\affiliation{Department of Astronomy/Steward Observatory, University of Arizona, 933 North Cherry Avenue, Tucson, AZ 85721, USA}

\date{\today}

\begin{abstract}
We explore the science prospects of a 14,000 deg$^2$ Kinematic Lensing (KL) survey with the Dark Energy Spectroscopic Instrument (DESI) and overlapping imaging surveys. KL infers the cosmic shear signal by jointly forward modeling the observed photometric image and velocity field of a disk galaxy. The latter can be constrained by placing multiple DESI fibers along the galaxy's major and minor axis, a concept similar to the DESI Peculiar Velocity Survey. We study multiple subset galaxy samples of the DESI Legacy Survey Data Release 9 catalog and quantify the residual shape noise, $\sigma_\epsilon$, of each sample as a function of cuts in $r$-band magnitude using mock observations. We conduct simulated likelihood analyses for these galaxy samples and find that a DESI-KL program can place highly interesting constraints on structure formation at very low redshifts, i.e. $\sigma_8(z<0.15)$. We conclude that if the $S_8$ tension consolidates as a phenomenon, a DESI-KL survey can provide unique insights into this phenomenon in the very late-time Universe. Given the different footprints of DESI and Rubin Observatory's Legacy Survey of Space and Time (LSST), lensing results from both surveys are highly complementary and can be combined into a joint lensing survey. We further note that DESI-KL benefits multiple additional science cases, e.g. studies of modified gravity models when combined with peculiar velocity surveys, and dark matter studies that are based on galaxy-galaxy lensing of dwarf galaxies.

\end{abstract}

\maketitle

\section{Introduction}
\label{sec:intro}
Weak gravitational lensing (WL) by the large-scale structure, so-called cosmic shear, is one of the core measurement techniques of future cosmological surveys such as Rubin Observatory's Legacy Survey of Space and Time \citep[LSST\footnote{\url{https://rubinobservatory.org/}}][]{DESC_SRD}, the Nancy Grace Roman Space Telescope \citep[Roman\footnote{\url{https://roman.gsfc.nasa.gov/}}][]{Roman_AFTA,HLIS-Multiprobe,HLIS-LSST}, and the Euclid\footnote{\url{https://www.esa.int/Science_Exploration/Space_Science/Euclid}}~\citep{Euclid} mission. While the community is developing a variety of analytical, numerical, and observational techniques to mitigate systematic effects that impact weak lensing, the options to reduce statistical uncertainties are limited. 

These statistical uncertainties are encoded in the cosmic shear covariance and can be broadly separated into shot noise, cosmic variance, and super-sample variance terms. The latter two can only be decreased by increasing the survey volume. The first, however, depends on the ratio of galaxy shape noise (squared) and the surface number density of galaxies $\sigma_\epsilon^2/n_\mathrm{gal}$. Commonly, the only way to reduce this noise component is to increase the galaxy number density of a survey by including fainter objects in the analysis. The shape noise $\sigma_\epsilon$ itself cannot be reduced further in imaging surveys since its main contribution, the unknown intrinsic ellipticity of the source galaxies, is a consequence of the degeneracy between the intrinsic ellipticity and the cosmic shear effect itself.

Recently, Kinematic Lensing (KL) has been proposed as a potential avenue to reduce shape noise in the case of disk galaxies~\citep{Blain02,Morales06,HKE+13,DSM_15a,DSM_15b,GTF20,GTF21,DiGiorgio21,XEH+23,PKH+23,HKX+24}. The core idea of KL is to combine the photometrically measured galaxy ellipticity with the spectroscopically measured galaxy velocity field. Since the velocity field and the photometric galaxy image transform differently under cosmic shear, their joint measurement tightly controls residual uncertainties in the galaxy inclination angle and hence significantly reduces the shape noise term. 

Other advantages of the KL method are: 1) the spectroscopic information eliminates photo-$z$ uncertainties, 2) the targeted galaxies have significantly higher photometric signal-to-noise ratios (SNR) compared to standard WL sample, which renders shape measurement uncertainties insignificant, and 3) since KL controls for the intrinsic galaxy orientation, it is immune to intrinsic galaxy alignment uncertainties. 

The main challenge of KL is to spectroscopically measure the velocity field for each of its source galaxies, a requirement that results in a significantly reduced galaxy sample size that is likely located at lower redshift compared to source samples used in traditional WL. 
Highly multiplexed spectroscopy from ground--based surveys, e.g. the ongoing Dark Energy Spectroscopic Instrument \citep[DESI\footnote{\url{https://www.desi.lbl.gov/}}][]{DESI1,DESI2} survey, the 4-metre Multi-Object Spectroscopic Telescope \citep[4MOST\footnote{\url{https://www.4most.eu/cms/home/}}][]{4MOST} survey, the MUltiplexed Survey Telescope \citep[MUST\footnote{\url{https://must.astro.tsinghua.edu.cn/en}}][]{MUST}, or high-resolution wide-field grism surveys from space, e.g. with the Roman Space Telescope (perhaps even with Euclid), are near-term opportunities to conduct KL measurements at scale. 

In this paper, we explore the potential cosmological science return of a KL survey using DESI fiber spectroscopy. Located at Kitt Peak National Observatory inside the 4-meter Mayall telescope, DESI observations are taken via 5000 robotic fibers that cover a 7.45 square degree field of view. The instrument measures galaxy spectra at a wavelength range of $360 < \lambda < 980$ nm with a spectral resolution of  2000--5500 depending on wavelength. Its primary science goal is to explore cosmic acceleration by studying the 3D clustering of galaxies with spectroscopic measurements, which requires one successful spectroscopic measurement per galaxy. 

A DESI-KL survey would rely on the idea that multiple fibers are placed across the galaxy in order to constrain the velocity field. This concept has been implemented successfully already as part of a DESI secondary targeting program, specifically the DESI Peculiar Velocity (PV) Survey \citep{DESIPV23}. The main difference between the DESI PV and a potential DESI-KL survey is that the latter would require a larger galaxy sample that would push to smaller/fainter objects and go to higher redshifts.

We have identified clustering at very low redshift, $\sigma_8(z<0.15)$, as a potentially high-profile science case for a DESI-KL survey. The core idea is that current results on $S_8$ from primary CMB measurements show a 2--3$\sigma$ $S_8$ tension with galaxy lensing and clustering surveys, while they are consistent with CMB lensing \cite{XEM+23,ACT_DR6_CMBL,FKM+23}. This gives rise to the idea that the $S_8$ tension may be most pronounced at low-$z$. Over the coming years, LSST, Euclid, and Roman will provide excellent $S_8$ measurements at intermediate redshifts while CMB lensing experiments such as Simons Observatory~\citep[SO\footnote{\url{https://simonsobservatory.org/}}][]{SO} will tightly constrain $S_8$ at higher redshifts. A DESI-KL survey is uniquely suited to provide insights into matter clustering at low-$z$ due to its increased shear signal-to-noise on a per-galaxy basis and the general advantage of lensing--based measurements being independent of any assumption between dark and luminous matter.

Beyond the science case of matter clustering at low redshift, we would like to stress that other science cases based on lensing will benefit greatly from KL. For example, dark matter models from stacked lensing of dwarf galaxies~\citep{LSL+20,LLG+24,TAW+23} and modified gravity studies that combine KL and peculiar velocities at very low redshifts~\citep{PPMH15,BHL20,NHW23}. While studying these cases in detail is beyond the scope of this paper they are nevertheless interesting to map out in the future.

We start this paper by explaining the basics of KL measurements and cosmological modeling (Sec. \ref{sec:KLbasic}). In Sec. \ref{sec:galsamples} we describe potential KL galaxy samples starting from the DESI Legacy Survey~\citep{LS_overview} and then considering subsets thereof, in particular the Bright Galaxy Survey sample~\citep{BGS_selection1,BGS_selection2}. Our simulation pipeline that infers the KL signal from simulated DESI imaging and spectroscopy is detailed in Sec. \ref{sec:pipe}. From the inferred KL signal we derive the residual shape noise contribution to the statistical uncertainties and then simulate a cosmological likelihood analysis for the various galaxy samples in Sec. \ref{sec:results}. We conclude in Sec. \ref{sec:conc}.

\section{Kinematic Lensing}
\label{sec:KLbasic}
We only briefly summarize the basics of Kinematic Lensing and refer the reader to \cite{XEH+23,PKH+23} for more details of our implementation. Throughout this work, we refer to $g_1$, $g_2$ as the reduced shear defined in equatorial coordinate system and $g_+$, $g_\times$ as the reduced shear defined in the galaxy frame where the X and Y axes are aligned with the photometric major and minor axes of the galaxy. 

\subsection{KL measurement basics}
\label{sec:esti}
The observed ellipticity $\hat{\bm{\epsilon}}\equiv\hat{\epsilon}_1+i\hat{\epsilon_2}$ of a galaxy can be expressed as a combination of its intrinsic shape $\hat{\bm{\epsilon}}^\mathrm{int}$ and the reduced shear $\bm{g}\equiv g_1+ig_2$ as~\citep{BS01_review}
\begin{equation}
\label{eqn:shear_transform}
    \hat{\bm{\epsilon}} = \frac{\hat{\bm{\epsilon}}^\mrm{int}+\bm{g}}{1+\bm{g}^*\hat{\bm{\epsilon}}^\mathrm{int}}.
\end{equation}
The standard WL industry is built on the assumption that $\hat{\bm{\epsilon}}^\mathrm{int}$ is (approximately) independently and randomly distributed. Thus, when measuring the two-point statistics over a large galaxy sample, the intrinsic shape averages to zero, and only the cosmic shear-induced correlation remains. Residual correlations of $\langle\hat{\bm{\epsilon}}^\mathrm{int}(\theta^\prime)\hat{\bm{\epsilon}}^\mathrm{int}(\theta^\prime+\theta)\rangle$ and $\langle\hat{\bm{\epsilon}}^\mathrm{int}(\theta^\prime)\bm{g}(\theta^\prime+\theta)\rangle$ are attributed to intrinsic alignment effects. Since $\hat{\bm{\epsilon}}^\mathrm{int}$ degenerates with $\bm{g}$ in Eq.~(\ref{eqn:shear_transform}), the measured two-point statistics of $\bm{g}$ has an \textit{irreducible} uncertainty $\propto{\sigma^2(\hat{\bm{\epsilon}}^\mathrm{int})}\approx 0.37^2$, aka the shape noise, which is much larger than typical shear signal $\sim 0.01$.

The observed line-of-sight (LoS) kinematic field of a galaxy $v_\mathrm{LoS}^\mathrm{obs}(\bm{\theta})$ is also distorted by cosmic shear as
\begin{equation}
    v_\mathrm{LoS}^\mathrm{obs}(\bm{\theta}) = v_\mathrm{LoS}(\textbf{A}\cdot\bm{\theta}),
\end{equation}
where $v_\mathrm{LoS}$ is the true LoS kinematic field and $\textbf{A}$ describes the shear distortion from observed position $\bm{\theta}_\mathrm{O}$ to the true position in the source plane $\bm{\theta}_\mathrm{S}$
\begin{equation}
\label{eqn:shear_mat}
     \textbf{A}\equiv \frac{\partial\bm{\theta}_{\mathrm{S}}}{\partial\bm{\theta}_{\mathrm{O}}}=
    (1-\kappa)\begin{pmatrix}
    1-g_1 & -g_2 \\
    -g_2 & 1+g_1
    \end{pmatrix}.
\end{equation}
For simplicity, we assume that the galaxy's major and minor axes are aligned with the X and Y axes and therefore $g_1$ and $g_2$ reduce to $g_+$ and $g_\times$.
As an example, for a disk galaxy with a $V_\mathrm{circ}$ rotation velocity plateau, the observed maximum rotation velocities along the major and minor axes of the galaxy are (neglecting terms higher than or equal to $\bigo{g^2}$)
\begin{equation}
\label{eqn:v_measured}
    \begin{aligned}
    v^\prime_\mathrm{major}&=-V_\mathrm{circ}\,\sin{i},\\
    v^\prime_\mathrm{minor}&= 
    V_\mathrm{circ}\,\sin{i}\,\cos{i}\,\left(1+\frac{1+e_\mathrm{obs}^2}{2e_{\mathrm{int}}}\right)g_\times\,,
    \end{aligned}
\end{equation}
where $i$ is the inclination angle of the galaxy and $e\equiv\left|\hat{\bm{\epsilon}}\right|$ is the ellipticity of the galaxy. For a disk galaxy of aspect ratio $q_z$, $e_\mathrm{int}$ can be related to $i$ by
\begin{equation}
\label{eqn:eint_geo}
    e_{\mrm{int}}=\frac{1-\sqrt{1-(1-q_z^2)\sin{^2i}}}{1+\sqrt{1-(1-q_z^2)\sin{^2i}}}\,.
\end{equation}
Also, disk galaxies follow the Tully-Fisher relation~\citep[TFR, ][]{TF1977,R11}
\begin{equation}
\label{eqn:TFR}
    \mrm{log}_\mathrm{10}(V_{\mrm{circ}})=a + b\, (M_\mrm{B}-M_\mrm{p})+\epsilon_\mathrm{TF}\,, 
\end{equation}
where $a$, $b$ are intercept and slope, $M_\mathrm{B}$ is the broad-band magnitude, $M_\mathrm{p}$ is the pivot magnitude, and $\epsilon_\mathrm{TF}$ is the intrinsic scatter of the TFR with a standard deviation of $\sigma_\mathrm{TF}$. The values $a$, $b$, and $\epsilon_\mathrm{TF}$ can be self-calibrated from the galaxy sample.

Naively, combining Eqs.~(\ref{eqn:shear_transform}-\ref{eqn:TFR}) we can solve for $g_+$ and $g_\times$
\begin{equation}
\label{eqn:reduced_shears}
    \begin{aligned}
    \hat{g}_+ &= \frac{e_\mrm{obs}^2-e_\mrm{int}^2}{2e_\mrm{obs}^2(1-e_\mrm{int}^2)},\\
    \hat{g}_\times&= \left|\frac{v_\mrm{minor}^\prime}{v_\mrm{major}^\prime}\right|\frac{2e_\mrm{int}}{\cos{i}\,(2e_\mrm{int}+1+e_\mrm{obs}^2)}\,,
    \end{aligned}
\end{equation}
given $e_\mrm{obs}$ measured from photometry images, $v_\mrm{minor}^\prime$ and $v_\mrm{major}^\prime$ measured from spectra. The intrinsic ellipticity and inclination angle can be indirectly inferred from TFR (broad-band magnitude) and the spectra. Intuitively speaking, cosmic shear impacts the galaxy photometry and LoS kinematic in different manners such that combining the two can \textit{break the degeneracy} between $\hat{\bm{\epsilon}}^\mathrm{int}$ and $\bm{g}$ in Eq.~(\ref{eqn:shear_transform}), and suppress the shape noise by an order of magnitude~\citep{HKE+13,XEH+23,PKH+23}.

The above equations provide a useful intuitive explanation of the KL concept. However, the practical implementation of a KL shear estimation involves building a full forward modeling pipeline of the sheared galaxy image and velocity field as a function of shear and the specific galaxy and instrument properties (see Sec. \ref{sec:pipe}).

\subsection{KL cosmology inference basics}
\label{sec:infere}
Similar to standard cosmic shear analyses of photometric galaxy surveys, we quantify the cosmological information content using second-order summary statistics of the shear field, e.g. the cosmic shear tomographic angular power spectrum, correlating galaxy samples in redshift tomography bins $i$ and $j$ assuming the Limber approximation:
\begin{equation}
    C_{\kappa\kappa}^{ij}(\ell)=\int \dd\chi \frac{q_\kappa^{i}(\chi)q_\kappa^{j}(\chi)}{\chi^2}P_{\delta\delta}(\ell/f_{K}(\chi),\,z(\chi)),
\end{equation}
where $\chi$ is the comoving distance, $f_K(\chi)$ is the comoving angular diameter distance, $\mathbf{\ell}$ is the wave vector, and $P_{\delta\delta}$ is the 3D nonlinear matter power spectrum, which is provided by \textsc{Halofit} in this work~\citep{SPJ+03,TSN+12}. The lensing efficiency for the $i$th bin is given by 
\begin{equation}
    q_\kappa^{i}(\chi)=\frac{3H_0^2\Omega_\mrm{m}}{2c^2}\frac{\chi}{a(\chi)}\int_\chi^{\chi_h}\dd \chi'\frac{n_\mrm{src}^{i}(\chi')}{\bar{n}_\mrm{src}^{i}}\frac{f_K(\chi'-\chi)}{f_K(\chi')}\,,
\end{equation}
where $n_\mrm{src}^{i}(\chi)\equiv n_\mrm{src}^{i}(z(\chi))\,\mathrm{d}z/\mathrm{d}\chi$ represents the redshift distribution of source galaxies in the $i$th bin, with $\bar{n}_\mrm{src}^{i}$ the 2D galaxy density within the same redshift bin. $\chi_h$ is the comoving distance to the horizon.

The corresponding covariance for these power spectra can be written as the sum of Gaussian, Non-Gaussian, and Super Sample Covariance terms~\citep{tj09,TH13,KE17}
\begin{equation}
\mat{C} = \mat{C}_{\rm{G}} + \mat{C}_{\rm{NG}} + \mat{C}_{\rm{SSC}}\,.
\end{equation}
The first term $\mat{C}_{\mrm{G}}$ reads
\begin{equation}
\label{eqn:C_G}
\begin{aligned}
   \mat{C}_{\mrm{G}} (\bm{C}_{\kappa\kappa}^{ij}(\ell_1) ,\,\bm{C}_{\kappa\kappa}^{kl}(\ell_2))=&
   \frac{4\pi\delta_{\ell_1\ell_2}}{\Omega_\mathrm{s}(2\ell_1+1)\Delta\ell_1}[\bar{\bm{C}}_{\kappa\kappa}^{ik}(\ell_1) \bar{\bm{C}}_{\kappa\kappa}^{jl}(\ell_1)\\
   &+\bar{\bm{C}}_{\kappa\kappa}^{il}(\ell_1)\bar{\bm{C}}_{\kappa\kappa}^{jk}(\ell_1)]\,,
\end{aligned}
\end{equation}
where $\Delta\ell_1$ is the $\ell_1$ bin width, $\Omega_\mathrm{s}$ is the survey area, and $\delta_{\ell_1\ell_2}$ is the Kronecker delta function. We refer to~\cite{KE17} and references therein for details of the modeling of the other two terms $\mat{C}_{\rm{NG}}$ and $\mat{C}_{\rm{SSC}}$.

The main difference between KL and traditional WL is encapsulated in the shape noise term $\sigma_{\bm{\epsilon}}$ which enters the Gaussian covariance via the definition
\begin{equation}
\bar{\bm{C}}_{\kappa\kappa}^{ij}(\ell_1)\equiv
    \bm{C}_{\kappa\kappa}^{ij}(\ell_1)+\delta_{ij}\frac{\sigma^2_{\bm{\epsilon}}}{\bar{n}^{i}_\mrm{src}}.
\end{equation}
The shape noise term, which in the above definition accounts for the shape noise of both shear components, is generally much smaller for KL compared to traditional WL, which has $\sigma_\epsilon\approx0.37$~\citep{Chang13}. We study the shape noise of KL samples in Sec. \ref{sec:pipe}.

These expressions for shear power spectra and their covariances allow us to run simulated likelihood analysis for specific DESI-KL samples in Sec. \ref{sec:results}. For more details of the modeling, see~\citep{EKS+14,KE17,XEH+23}.

\section{DESI galaxy samples}
\label{sec:galsamples}
\begin{figure*}
    \includegraphics[width=\linewidth]{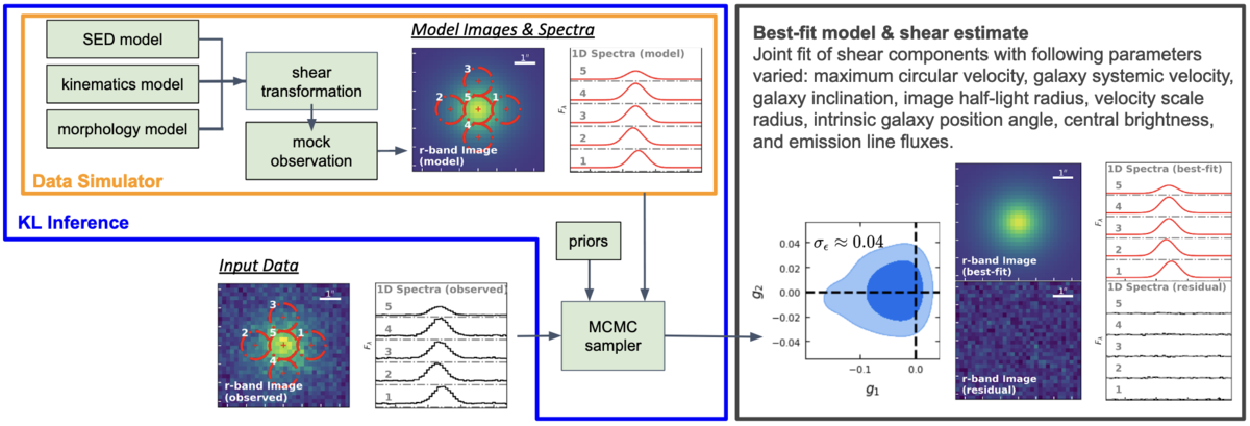}
    \caption{Architecture of the KL shear inference pipeline: the orange box (top) illustrates the data simulator, which forward models images and spectra from an astrophysical model, input shear, and instrument characteristics. We show five illustrative 1D spectra which are taken by the fibers annotated by the red circular apertures on the mock $r$-band image. The diameter of the red circular apertures corresponds to the DESI fiber diameter $1.\arcsec5$. The blue box illustrates the Bayesian inference stage, which produces best-fitting models and shear posterior (black box) for a set of input data, which is also generated by the data simulator in this work with realistic SNR.}
    \label{fig:pipe}
\end{figure*}

\begin{figure}
\centering
    \includegraphics[width=\linewidth]{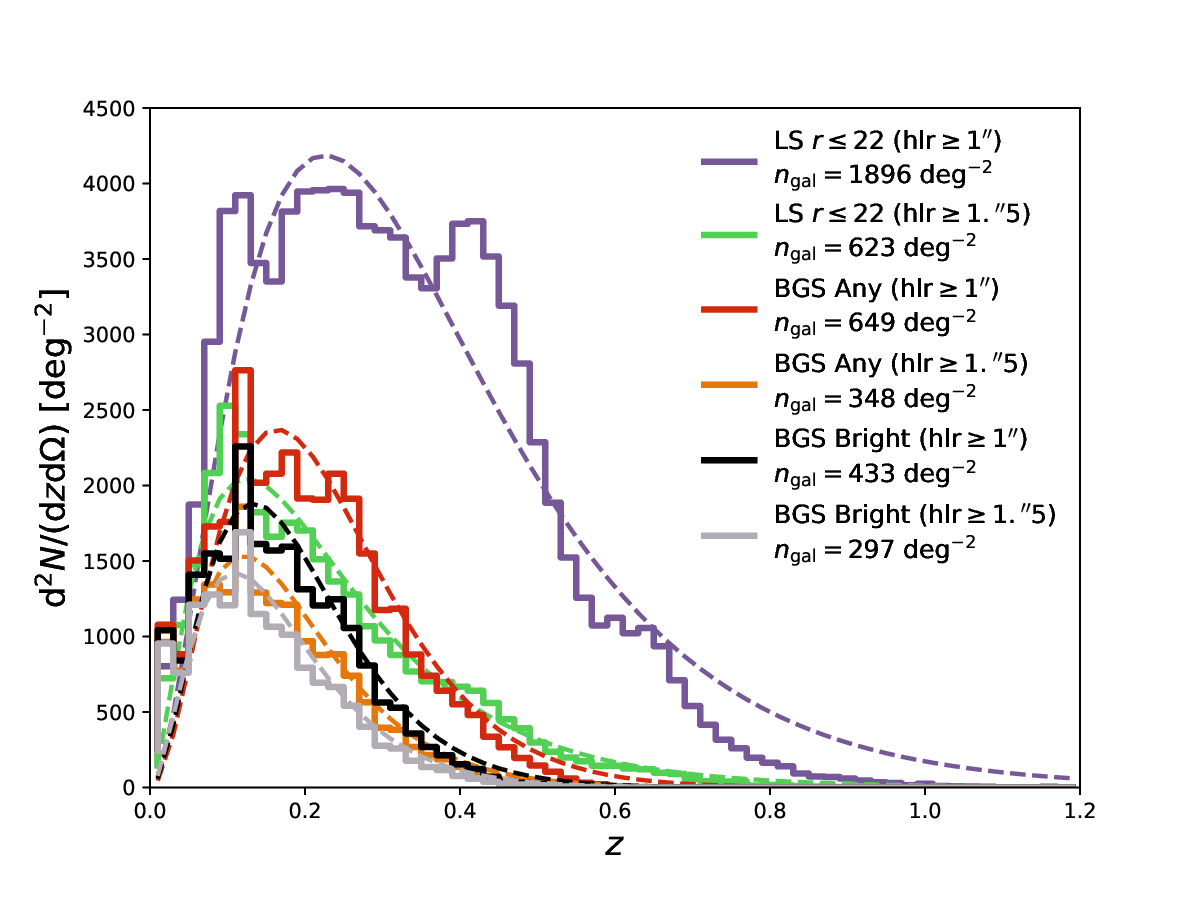}
    \caption{The redshift distribution of the KL samples defined in Sec.~\ref{sec:galsamples}. The samples mentioned in the figure legend, from top to bottom, correspond to the samples 1 to 6 defined in Table~\ref{tab:sample_def}. The redshift of samples 1 and 2 are estimated from their median photo-$z$ in the tractor catalog, while the redshift of samples 3-6 are spectroscopic redshift measured in SV3. We show the parametric fits to those redshift distributions in dashed lines with the same color.}
    \label{fig:zdist}
\end{figure}

\begin{table}
    \centering
    \begin{tabular}{ccccc}
    \toprule
        Sample & \shortstack{Selection\\Criteria} & $n_\mathrm{gal}$ (arcmin$^{-2}$) & $z_0$ & $\alpha$\\
        \tableline 
        1 & 1.a \& 2.a & 0.527 & 0.0894 & 0.8826 \\
        2 & 1.a \& 2.b & 0.173 & 0.0259 & 0.6835 \\
        3 & 1.b \& 2.a & 0.180 & 0.1027 & 1.1616 \\
        4 & 1.b \& 2.b & 0.097 & 0.0569 & 0.9701 \\
        5 & 1.c \& 2.a & 0.120 & 0.0777 & 1.1031 \\
        6 & 1.c \& 2.b & 0.083 & 0.0504 & 0.9537 \\
        \tableline
    \end{tabular}
    \caption{Summary of KL sample target selection criteria and the resulting sample number density and redshift distribution. We define six samples with their target selection criteria shown in the second column and the surface number density in the third column. We show the best-fitting parameters of $n(z)$ distribution in Eq.~(\ref{eqn:nz_param}) in columns four and five.}
    \label{tab:sample_def}
\end{table}

The basic idea of a DESI-KL survey is to place multiple fibers across the galaxy image in order to constrain the velocity field. This concept has already been implemented successfully for the DESI peculiar velocity survey (DESI-PV) \citep{DESIPV23}.
The goal of the DESI PV survey is to measure the two-point statistics between galaxy overdensity and peculiar velocity at $z\leq0.1$ to constrain the late-time structure growth $f\sigma_8$. 
The DESI PV sample mainly consists of early-type galaxies (ETGs) and late-type galaxies (LTGs). Both redshift and distance are needed to calculate peculiar velocities. Distances to the galaxies are estimated via the fundamental plane (FP) for the ETGs and the Tully-Fisher relation for the LTGs. To measure the rotation velocity of LTGs, two offset fibers are placed along the major axis with $\pm 0.4R_{26}$ from the galaxy center, in addition to the central fiber observation conducted as part of the BGS program~\citep[see the Fig. 8 in][for an example measurement]{DESIPV23}. Here $R_{26}$ is the semimajor axis radius measured at $\mu=26$ mag arcsec$^{-2}$ $r$-band isophote. The final DESI PV survey is predicted to measure PV from $\sim 133,000$ FP--based and $\sim53,000$ TF--based galaxies over the $14,000$ deg$^2$ DESI 5-year footprint and expect a $4$ percent precision on $f\sigma_8(z<0.15)$.  

The DESI-KL sample requirements differ from that of the PV sample in that the KL sample requires a significantly higher galaxy number density which also includes galaxies at higher redshifts. Consequently, we include smaller galaxies (half-light radius $>1.^{\prime\prime}0$, abbreviated as hlr thereafter) than the PV sample and we forward model the velocity field. An example data set is shown in Fig.~\ref{fig:pipe}, where four offset fibers are placed along both the major and minor axes of the galaxy, in addition to the central fiber observed in BGS.

We start with the DESI Legacy Imaging Survey \citep{LS_overview} as the basis for our galaxy sample selection. The Legacy Survey covers $14,000$ deg$^{2}$ spanning approximately in declination $-18\degree< \delta < +84\degree$. The Legacy Survey Data Release 9\footnote{\url{https://www.legacysurvey.org/dr9/}}~\citep[LS DR9 from hereon][]{SDH+21} contains three optical bands in $(g,\,r,\,z)$ and four infrared bands from WISE. The optical bands images are composed by three survey projects: the Beijing-Arizona Sky Survey~\citep[BASS,][]{BASS}, the DECam Legacy Survey~\citep[DECaLS,][]{LS_overview}, and the Mayall $z$-band Legacy Survey~\citep[MzLS,][]{LS_overview}. We use the tractor sweep catalog ~\citep{LHM16,L20} for the photometry, shape, photo-$z$, and morphology measurements. 

From the LS DR 9 catalog, we select disk galaxies with half-light radii larger than $1\arcsec$ to accommodate offset fibers in the galaxy. We start from a modified version of the LTGs target selection criterion in~\citep{DESIPV23} since both the KL and the PV samples are targeted at galaxies following the TFR, and the LTGs target selection in \citep{DESIPV23} has a high precision for LTGs (85.1--92.7 percent of the selected targets are classified as LTGs). We identify six possible galaxy samples suitable for a DESI-KL survey (see Fig. \ref{fig:zdist}) which are described below (see Appendix~\ref{sec:append_ts} for the exact target selections): 

We first select bright galaxies from the tractor catalog. Following~\citep{BGS_selection2}, three parallel selection criteria are proposed, ranked by galaxy number density from high to low:
\begin{itemize}
    \item Criterion 1.a LS $r\leq22$: This is the most relaxed selection criterion among the three. We pose similar selection criteria as the one in~\citep{BGS_selection2} but relax the $r$-band magnitude cut to 22. The motivation is to include more galaxies to study the trade-off between number density v.s. SNR.  
    \item Criterion 1.b BGS Any: This is more stringent than 1.a and we include all galaxies that are either selected as BGS Bright or BGS Faint. The benefit is that their redshift and emission line properties are already measured by BGS, which eliminates many of the uncertainties in target selection using the photometric catalog. Also, we only need to take offset fiber exposures for the BGS Any sample. 
    \item Criterion 1.c BGS Bright: This criterion selects galaxies from the BGS Bright sample. This ensures high SNR but reduces the number density.
\end{itemize}
We then select disk galaxies by imposing the following two parallel criteria
\begin{itemize}
    \item Criterion 2.a: The tractor morphology is classified as inclined exponential, round exponential, or Sersic profile and with Sersic index $\leq 2$, and the above-atmosphere hlr $\geq 1\arcsec$.
    \item Criterion 2.b: Same as 2.a but with a more conservative cut on the above-atmosphere hlr $\geq 1.\arcsec5$.
\end{itemize}
Altogether we get six galaxy samples that satisfy one criterion within 1.a-c and one criterion in 2.a-b. We show the spectroscopic or median-photometric redshift distribution $\mathrm{d}^2N/(\mathrm{d}z\mathrm{d}\Omega)$ of these six samples in Fig.~\ref{fig:zdist} as binned histograms. We also fit a parametric redshift distribution 
\begin{equation}
    \label{eqn:nz_param}
    n(z) \propto z^2 \mathrm{exp}[-(z/z_0)^\alpha]\,, 
\end{equation}
to the histogram of each sample, and show the corresponding best-fitting $n(z)$ in dashed lines. The smoothed parametric $n(z)$ is then used in the cosmological forecast in Sec.~\ref{sec:results}. We summarize the key sample properties related to kinematic lensing analysis in Table~\ref{tab:sample_def}. These samples span a sufficiently large variety for us to consider. 
For convenience, we refer to samples 1 and 2 as ``the Legacy Survey samples'' and use them interchangeably in the following sections due to their nature as photometric samples. If not mentioned explicitly, we mean sample 1 by ``the Legacy Survey sample'' since it is much more constraining than sample 2. 

\section{Simulating KL with a DESI galaxy sample}
\label{sec:pipe}

\subsection{KL simulation overview}
\label{sec:kltools_overview}

We use a modified version of the KL simulation pipeline presented in \citep{PKH+23} to quantify the shape noise level for a realistic DESI galaxy sample. We illustrate the simulation pipeline framework in Fig.~\ref{fig:pipe} and summarize the key steps below.

Given observed 2D broad-band images and 1D fiber spectra (denoted as data vector $\bm{D}$), we forward model both images and spectra (denoted as model vector $\bm{M}$) to derive the maximum a posteriori probability (MAP) estimate of model parameters $\bm{\theta}$. The posterior is computed as
\begin{equation}
    p(\bm{\theta}|\bm{D})\propto L(\bm{D}|\bm{\theta})p(\bm{\theta}),
\end{equation}
where $p(\bm{\theta})$ is the prior of model parameters.
We assume Gaussian likelihood for both the 2D images and the 1D spectra and ignore noise correlation
\begin{equation}
    L(\bm{D}|\bm{\theta}) = \prod_{i=1}^{N}\mathrm{exp}\left(-\frac{\left(\bm{D}_i-\bm{M}_i(\bm{\theta})\right)^2}{2\sigma_i^2}\right)\,,
\end{equation}
where the subscript $i\in[1,\,N]$ stands for the $i$th image/spectrum observation and $\sigma_i$ is the standard deviation of the noise in the $i$th observation.

In a real KL measurement, $\bm{D}$ comes from observed data. In this forecast work, however, we use the same model to both generate mock data at fiducial parameter $\bm{\theta}_\mathrm{fid}$ and fit the MAP parameter to recover $\bm{\theta}_\mathrm{fid}$. 

The KL measurement pipeline contains a data simulator to forward model mock images and spectra given input parameters $\bm{\theta}$. The data simulator first generates an intensity profile $I(\bm{x}|\bm{\theta})$ (dimensionless, above the atmosphere) and line-of-sight (LoS) rotation velocity field $v_\mathrm{LoS}(\bm{x}|\bm{\theta})$ ($\mathrm{km}\,\mathrm{s}^{-1}$) that are distorted by cosmic shear, as well as an observer-frame spectral energy distribution (SED) model $F_\lambda^\mathrm{obs}(\lambda|\bm{\theta})$ ($\mathrm{erg}\,\mathrm{s}^{-1}\,\mathrm{cm}^{-2}\,\mathrm{nm}^{-1}$). Then it constructs the 3D photon distribution
\begin{equation}
    f(\bm{x},\lambda|\bm{\theta})=I(\bm{x}|\bm{\theta})\, F_\lambda^\mathrm{obs}\left(\left.\frac{\lambda}{1+v_\mathrm{LoS}(\bm{x})/c}\right|\bm{\theta}\right)\frac{\lambda}{hc}\,.
\end{equation}
We use $f(\bm{x},\lambda|\bm{\theta})$ to further generate mock 2D photometry image
\begin{equation}
\label{eqn:mock_img}
    \bm{M}_\mathrm{I}(\bm{x}|\bm{\theta})=\frac{At}{G}(\mathcal{P}\ast I)(\bm{x})\int\mathrm{d}\lambda \frac{F_\lambda^\mathrm{obs}(\lambda|\bm{\theta})}{hc/\lambda} \mathcal{T}(\lambda)\,,
\end{equation}
or mock 1D fiber spectrum
\begin{equation}
\label{eqn:mock_spec}
    \bm{M}_\mathrm{S}(\lambda|\bm{\theta}) = \frac{At}{G}\mathcal{R}\cdot\int\mathrm{d}^2\bm{x}f(\bm{x},\lambda^\prime|\bm{\theta})(\mathcal{M}\ast\mathcal{P})(\bm{x})\mathcal{T}(\lambda^\prime)\,,
\end{equation}
where $A$ is the effective collecting area of the telescope, $t$ is the exposure time, $G$ is the detector gain, $\mathcal{P}$ is the point-spread function (PSF), $\mathcal{T}(\lambda)$ is the system throughput (including atmosphere transmission, optical train throughput, and detector quantum efficiency), $\mathcal{M}$ is the fiber aperture mask, $\mathcal{R}$ is the resolution matrix of DESI spectrograph \citep{BS10,GBK+23}, and $\ast$ means convolution. We then sample the model space (see Table \ref{tab:kl_params}) and evaluate the cosmic shear uncertainties $\sigma_{g_1}$ and $\sigma_{g_2}$ from their 1D projected posteriors. These allow us to compute the shape noise as $\sigma_\epsilon=\sqrt{\sigma_{g_1}^2+\sigma_{g_2}^2}$. 

\subsection{Single galaxy measurement}
\label{sec:single_sn}

In order to generate the 3D photon distribution $f(\bm{x},\lambda|\bm{\theta})$, we adopt the following parametric models for $I(\bm{x}|\bm{\theta})$, $v_\mathrm{LoS}(\bm{x}|\bm{\theta})$, and $F_\lambda^\mathrm{obs}(\lambda|\bm{\theta})$: 

\paragraph{Morphology} We use the inclined exponential profile in \textsc{GalSim}~\citep{galsim15} to model the intensity profile. It is controlled by five parameters: the inclination angle $i$ ($i=0$ for face-on), the position angle $\theta_\mathrm{int}$ ($\theta_\mathrm{int}=0$ for alignment with the $x$-axis), the scale radius $R_s$, and reduced shear $g_1$, $g_2$. The 3D intensity distribution is determined as (in the galaxy intrinsic cylindrical coordinate)
\begin{equation}
    I_\mathrm{3D}(R,h|R_s)\propto \mathrm{sech}^2(h/h_s)\mathrm{exp}(-R/R_s),
\end{equation}
where $h$ is the distance to the $x$-$y$ plane, $R$ is the projected radius in the $x$-$y$ plane, and $h_s$ is the scale height. We assume a fixed aspect ratio $q_z\equiv h_s/R_s=0.1$ in our analysis. The 3D profile is then projected into 2D plane according to $i$ and $\theta_\mathrm{int}$ to get $I_{2D}(\bm{x}|i,\theta_\mathrm{int},R_s)$, and then distorted by $g_1$, $g_2$
\begin{equation}
    I(\bm{x}|g_1,g_2,i,\theta_\mathrm{int},R_s)=I_\mathrm{2D}(\textbf{A}\cdot\bm{x}|i,\theta_\mathrm{int},R_s)\,,
\end{equation}
where $\textbf{A}$ is the distortion matrix defined in Eq.~(\ref{eqn:shear_mat}).

\paragraph{Spectral Energy Distribution} The SED is separated into two components: continuum and emission lines. We adopt a typical continuum of barred spiral galaxies and add Gaussian emission lines on top of the continuum. The observer-frame continuum is controlled by two parameters: redshift $z$ and the flux density at observer-frame pivot wavelength $\lambda_\mathrm{norm}^\mathrm{obs}=850$ nm. Each emission line is controlled by two parameters: redshift $z$ and flux $F$. We assume a fixed rest-frame $1\sigma$ emission line width of 0.05 nm in this work, which corresponds to a velocity dispersion $\sim 23\,\mathrm{km}\,\mathrm{s}^{-1}$. We mainly consider four emission lines in this work: $\mathrm{H}\,\alpha$, $[\mathrm{O}\,\textsc{ii}]$ doublet, $[\mathrm{O}\,\textsc{iii}]4960$ and $[\mathrm{O}\,\textsc{iii}]5008$. 

\paragraph{Velocity field} For the kinematic structure, we assume an infinitesimal thin circular disk and the rotation velocity can be expressed as a regular arctan function~\citep{C97,GGG+14}
\begin{equation}
    v_\mathrm{rot}(R)=\frac{V_\mathrm{circ}}{\pi/2}\mathrm{tan}^{-1}(\frac{R}{R_\mathrm{vscale}})\,,
\end{equation}
where $R_\mathrm{vscale}$ is the scale radius of the rotation curve. We then project the rotation velocity along LoS based on $i$, $\theta_\mathrm{int}$ and add the peculiar velocity $v_0$, and distort the projected velocity field by $g_1$, $g_2$~\citep[see][for details on the projection and shear distortion]{PKH+23,XEH+23}. 

Integrating the models above, we are sampling 10 parameters during KL shear inference for a single emission line analysis with $\mathrm{H}\,\alpha$: $\bm{\theta}$=($g_1$, $g_2$, $\theta_\mathrm{int}$, $\mathrm{sin}\,i$, $v_0$, $V_\mathrm{circ}$, $R_\mathrm{vscale}$, $R_h$, $F_\mathrm{cont}$, $F_{\mathrm{H}\,\alpha}$). For multiple emission line analyses, we add three more parameters ($F_{[\mathrm{O}\,\textsc{ii}]}$, $F_{[\mathrm{O}\,\textsc{iii}]4960}$, $F_{[\mathrm{O}\,\textsc{iii}]5008}$). The priors of those parameters are summarized in Table~\ref{tab:kl_params}. Specifically, we include the TFR prior on $V_\mathrm{circ}$ as a Gaussian prior of 80 km s$^{-1}$, which corresponds to $\sigma_\mathrm{TF}\approx0.11$ dex for $V_\mathrm{circ}=300$ km s$^{-1}$. We note that this is a relatively conservative choice since $\sigma_\mathrm{TF}$ generally ranges from 0.05 to 0.12~\citep{CBB07,MBS+11,RMG+11}.

For single galaxy KL measurement, the data vector $\bm{D}$ is produced at some fiducial parameter $\bm{\theta}_\mathrm{fid}$ and we use \textsc{emcee}~\citep{emcee} to sample the posterior given the priors and realistic noise and use $\textsc{getdist}$~\citep{getdist} to derive the 1D marginalized $g_1$, $g_2$ uncertainties.

\begin{table*}
    \centering
    \begin{tabular}{llccl}
        \toprule
        Parameter & Description & Fiducial Value & Prior & Units\\
        \tableline
        $g_+$ & Reduced shear component & 0.0 & $\mathcal{U}$(-0.5, 0.5) & 1\\
        $g_\times$ & Reduced shear component & 0.0 & $\mathcal{U}$(-0.5, 0.5) & 1\\
        $\theta_\mathrm{int}$ & Intrinsic galaxy position angle & 0.0 & $\mathcal{U}$(-$\pi$, $\pi$) & rad\\
        $\mathrm{sin}\,i$ & Galaxy inclination & 0.05--0.95 & $\mathcal{U}$(0, 1) & 1\\
        $v_0$ & Galaxy systemic velocity & 0.0 & $\mathcal{N}$(0, 10) & km s$^{-1}$\\
        $V_\mathrm{circ}$ & Maximum rotation velocity & 300 & $\mathcal{N}$(300, 80) & km s$^{-1}$\\
        $R_\mathrm{vscale}$ & Velocity scale radius & 0.5--2.0 & $\mathcal{U}$(0.1, 5) & arcsec\\
        $R_h$ & Intensity half-light radius & 0.5--2.0 & $\mathcal{U}$(0.1, 0.5) & arcsec\\
        $F_\mathrm{cont}$ & Continuum flux density at 850 nm & 0.3--75.4 & $\mathcal{U}$($10^{-5}$, $10^{5}$) & $10^{-16}$ erg s$^{-1}$ cm$^{-2}$ nm$^{-1}$\\
        $F_{\mathrm{H}\,\alpha}$ & $\mathrm{H}\,\alpha$ flux  & 1.20--301.43 & $\mathcal{U}$($10^{-5}$, $10^{5}$) & $10^{-16}$ erg s$^{-1}$ cm$^{-2}$\\
        $F_{[\mathrm{O}\,\textsc{ii}]}$ & $[\mathrm{O}\,\textsc{ii}]$ flux & 0.88--221.05 & $\mathcal{U}$($10^{-5}$, $10^{5}$) & $10^{-16}$ erg s$^{-1}$ cm$^{-2}$\\
        $F_{[\mathrm{O}\,\textsc{iii}]4960}$ & $[\mathrm{O}\,\textsc{ii}]4960$ flux & 0.24--60.29 & $\mathcal{U}$($10^{-5}$, $10^{5}$) & $10^{-16}$ erg s$^{-1}$ cm$^{-2}$\\
        $F_{[\mathrm{O}\,\textsc{iii}]5008}$ & $[\mathrm{O}\,\textsc{ii}]5008$ flux  & 0.28--70.33 & $\mathcal{U}$($10^{-5}$, $10^{5}$) & $10^{-16}$ erg s$^{-1}$ cm$^{-2}$\\
        \tableline
        $q_z$ & Aspect ratio & 0.1 & fixed & 1\\
        $\sigma^\mathrm{int}_\lambda$ & Rest-frame emission line width & 0.05 & fixed & nm\\
        $z$ & Redshift & 0.3 & fixed & 1\\
        \tableline
        \end{tabular}
    \caption{A list of KL shear measurement model parameters. We show the fiducial values used for mock data $\bm{D}$ generation in the third column and the priors in the fourth column. Some parameters have a range of fiducial values when generating the mock data and we show their range in the fiducial value. We denote flat priors as $\mathcal{U}$(min, max) and Gaussian priors as $\mathcal{N}$(mean, std).}
    \label{tab:kl_params}
\end{table*}

\subsection{Mock observation configuration}
\label{sec:obs_config}

\subsubsection{Photometric images}
\label{sec:obs_image}

In our baseline forecast, the image component comes from the Legacy Survey, specifically the $g,r,z$ bands images in BASS, MzLS, and DECaLS. The median $5\sigma$ depth for an extended source with $0.45\arcsec$ hlr in those bands are $g=24.0$, $r=23.4$, and $z=22.5$ respectively. We note that this is the most conservative choice, and for example, UNIONS\footnote{\url{https://www.skysurvey.cc/}}~\citep{GKF+22}, LSST, or Euclid overlap would provide a much deeper imaging dataset.

We set the observation parameters $A$, $t$, $G$ in Eq.~(\ref{eqn:mock_img}) such that an object with 24.0/23.4/22.5 magnitude in $g/r/z$ band has a median SNR of 5. To study possible synergies with LSST, we also explore a range of $r$-band depth ($5\sigma$ depth of point source): 24.5, 25.81, 26.5, and 27.04, where the 25.81 and 27.04 correspond to LSST Y1 and Y10 depth. We assume a typical seeing of $1.\arcsec0$ and a pixel scale of $0.\arcsec2637$. 

\subsubsection{Fiber spectra}
\label{sec:obs_spec}

The fiber spectra simulation is performed assuming typical dark-time conditions. We adopt the dark sky model presented in \textsc{specsim} as the nominal dark condition, with airmass of 1, no extinction, and a seeing FWHM of $1.\arcsec 0$. The sky background $r$-band surface brightness is $r=20.61$ mags/arcsec$^2$. 

To infer the kinematic information, we need spectra from fibers offset from the galaxy center to sample different parts of $v_\mathrm{LoS}$. Positions of special interest are along the major and minor axes where large and zero central wavelength shifts are expected. We consider two fiber configurations in this work:
\begin{itemize}
    \item Five fibers configuration (major+minor): As shown in Fig.~\ref{fig:pipe}, we place four offset fibers along the major and minor axes, right next to the central fiber, which is taken from the main BGS program. We assume the same exposure time $t_\mathrm{nominal}$ for all four offset fibers and assume 180 seconds nominal exposure time for the central fiber.
    \item Three fibers configuration (semi-(major+minor)): Other than the central fiber, we place one fiber in the semimajor axis and another one in the semiminor axis (e.g. fiber 2, 3, 5 in Fig.~\ref{fig:pipe}). This will reduce the fibers needed while still covering the information along the major and minor axes. In order to compare the shape noise performance purely due to fiber placements, the exposure time is set to two times the nominal time used in the five-fiber cases. 
\end{itemize}
We consider two nominal offset-fiber exposure times, 600 and 900 seconds, in this analysis. The reduced 1D spectra are cropped into 5 nm wide snippets around the emission lines with a dispersion of 0.08 nm/pixel. We assume the DESI instrument settings in Eq.~(\ref{eqn:mock_spec}). We list the instrument properties used in this work in Table~\ref{tab:inst_properties}. 

\begin{table}
    \centering
    \begin{tabular}{lcc}
        \toprule
        \rule{0pt}{5ex}
        Properties & \shortstack{KPNO\\Mayall 4-meter} & \shortstack{CTIO\\Blanco 4-meter} \\
        \tableline
        Effective diameter (cm) & 332.42 & 378.29\\
        Pixel scale ($\arcsec/\mathrm{pix}$) & - & 0.2637 \\
        PSF FWHM ($\arcsec$)& 1.0 & 1.0\\
        Read noise (e$^{-}$/pix) & 3.41($b$)/2.6($r$)/2.6($z$)  & 7 \\
        Gain (e$^{-1}$/ADU) & 1 & 4 \\
        Dispersion (nm/pix) & 0.08  & -\\
        \tableline
    \end{tabular}
    \caption{Summary of instrument properties and observation conditions adopted in mock images and spectra generation. We show the properties of DESI (KPNO Mayall 4-meter telescope) in the second column and DECaLS (CTIO Blanco 4-meter telescope) in the third column.}
    \label{tab:inst_properties}
\end{table}

\begin{figure*}
    \includegraphics[width=\linewidth]{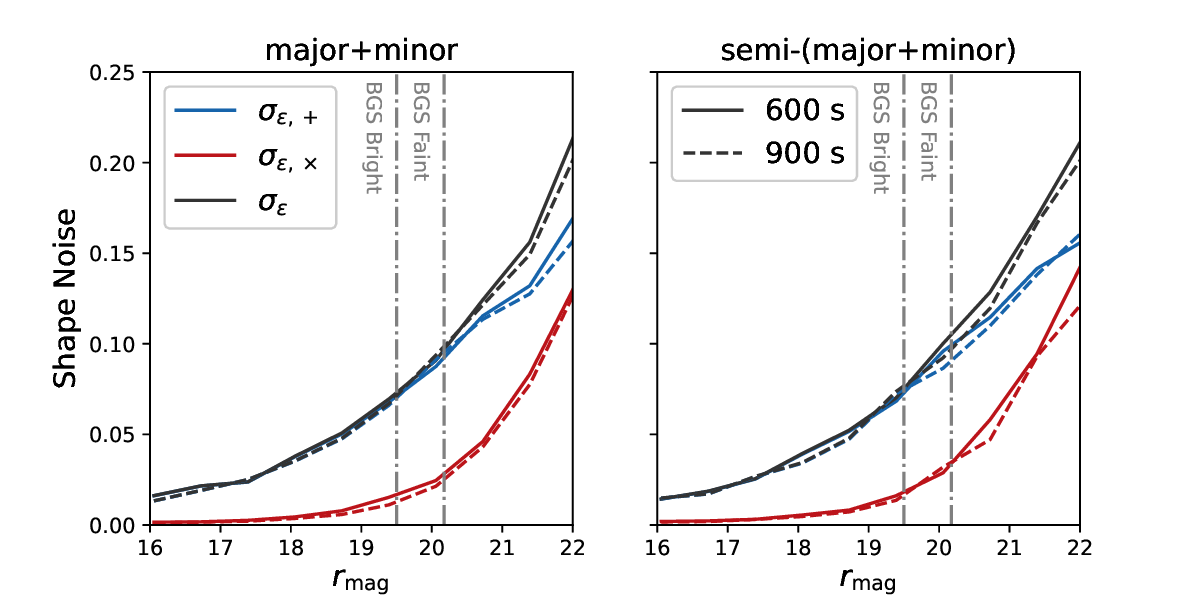}
    \caption{Shape noise as a function of $r$-band magnitude. The data set adopted here is $g$, $r$, $z$ band images at the depth of LS DR9, and $[\mathrm{O}\,\textsc{ii}]$, $[\mathrm{O}\,\textsc{iii}]4960$, $[\mathrm{O}\,\textsc{iii}]5008$, $\mathrm{H}\,\alpha$ emission lines taken at both the central fiber (from BGS main survey) and the offset fibers (assuming 600 or 900 seconds exposure). On the left panel, we show the results where four offset fibers are placed along major and minor axes, while on the right panel we show where two offset fibers are placed along the semi-major and semi-minor axes but with exposure time doubled. We show results with 600 (900) seconds nominal exposure time in solid (dashed) lines. The shape noises of the $g_+$, $g_\times$ components are shown in blue and red. The combined shape noise $\sigma_\epsilon$ is shown in black. We annotate the $r$-band limit for the BGS Bright and Faint samples with vertical dotted-dashed grey lines.}
    \label{fig:SN_rmag}
\end{figure*}

\begin{figure}
    \includegraphics[width=\linewidth]{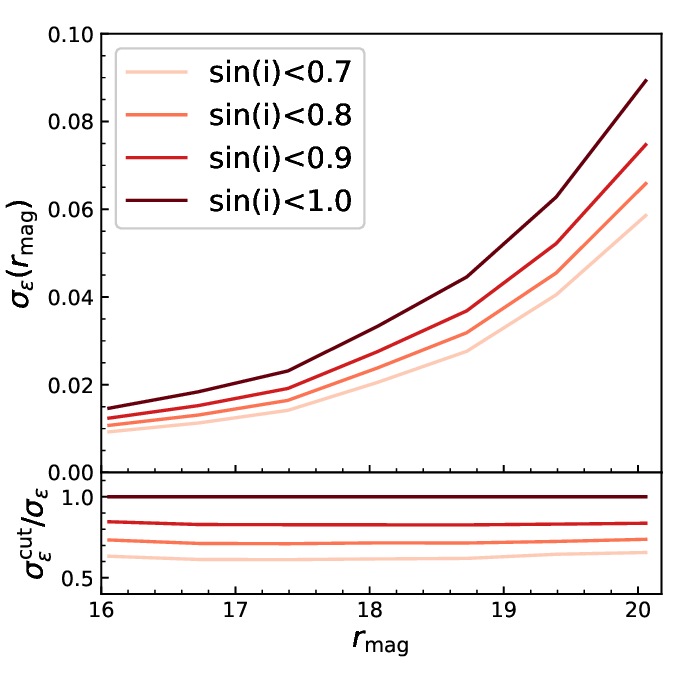}
    \caption{Changes in shape noise when selecting galaxies based on the observed inclination angle. We show the shape noise in the top panel and the ratio between shape noise with inclination cut and without in the bottom. While shape noise for lower inclination angle is generally lower, the loss in number density more than offsets the gain in shape noise. }
    \label{fig:SN_sinicut}
\end{figure}

\begin{figure*}
    \includegraphics[width=\linewidth]{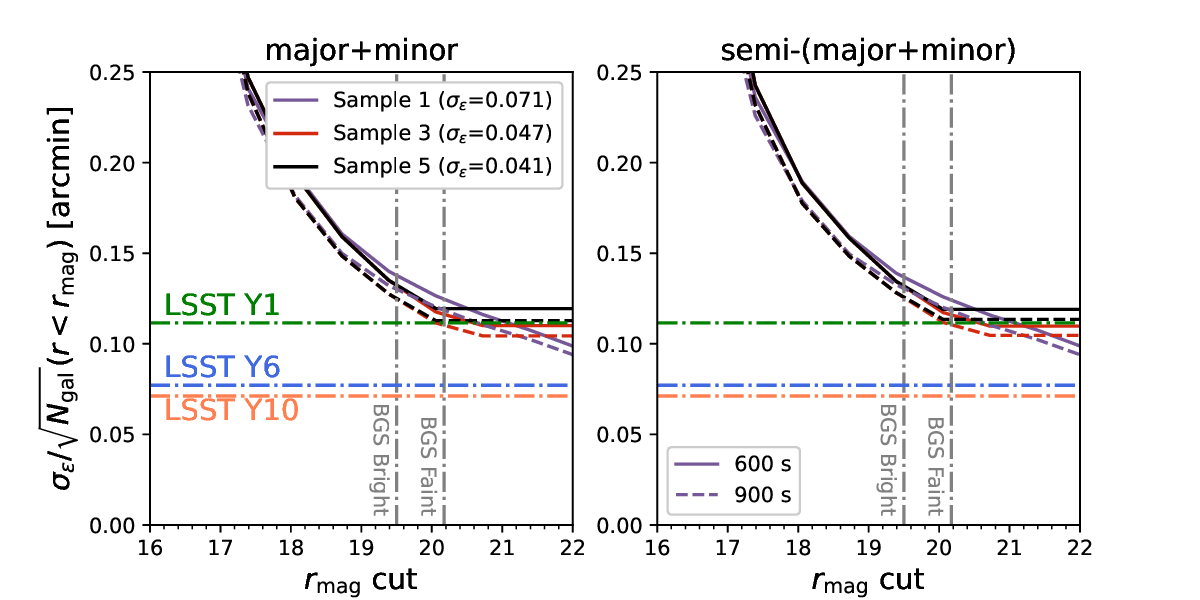}
    \caption{Combination of $\sigma_\epsilon/\sqrt{n_\mathrm{gal}}$ as a function of $r$-band magnitude cut for different KL samples. We show the results assuming four offset fibers on the left panel and two offset fibers on the right panel. Solid and dashed lines are results assuming 600 and 900 seconds nominal offset-fiber exposure time. We also annotate the full-sample $\sigma_\epsilon/\sqrt{n_\mathrm{gal}}$ of LSST Y1, Y6, and Y10 with the green, blue, and orange dotted-dashed lines. The curves of samples 3 and 5 are saturated after their faintest $r$ magnitude.}
    \label{fig:SN_over_ngal}
\end{figure*}

\subsection{Sample-averaged shape noise}
\label{sec:average_sn}

Shear uncertainty of kinematic lensing is a function of images and spectra SNR, galaxy size, and inclination angle~\citep{Chang13,PKH+23,XEH+23}. Bright, well-resolved, and low-inclination galaxies have lower shape noise. To evaluate the average shape noise for a galaxy sample, we evaluate $\sigma_\epsilon$ on a 3D grid of ($r_\mathrm{mag}$, $\mathrm{sin}i$, $R_h$), which covers the parameter space spanned by galaxy sampled defined in Sec.~\ref{sec:galsamples}. We generate mock data on each grid point and run our KL simulation pipeline as described in Sec.~\ref{sec:single_sn}. Then, we evaluate the average of $\sigma_\epsilon$ on the grid, weighted by galaxy number density and inverse variance of cosmic shear at each grid point. We quote the weighted average shape noise as the effective shape noise of the sample. 

We use the $r$-band magnitude and half-light radius distribution derived from LS DR9 catalog and assume a random distribution for the inclination angle. 
We first show the shape noise as a function of $r$-band magnitude and marginalized over inclination angle and hlr in Fig.~\ref{fig:SN_rmag}, which applies to all the KL samples. The dataset assumed in Fig.~\ref{fig:SN_rmag} includes $g$, $r$, $z$ band images at LS DR9 depth and five (left panel) or three (right panel) fiber spectra taken. Results for two nominal exposure times, 600 and 900 seconds, are shown in the figure, and we define the 600-second scenario as our baseline going forward. 

We also explore how shape noise depends on galaxy inclination by down-selecting the sample based on the observed inclination angle $\mathrm{sin}\,i < 0.7,\,0.8,\,0.9,\,$ and 1.0. The results are shown in Fig.~\ref{fig:SN_sinicut}. Although excluding high-inclination galaxies can improve the average shape noise, the loss in galaxy number density is too severe a hit for this approach to be promising. Nevertheless, if we need to down-select galaxies from a given population, e.g. due to a lack of fibers, the inclination angle is an important criterion.

Based on the relation between shape noise and $r$-band magnitude in Fig.~\ref{fig:SN_rmag}, we derive the weighted shape noise $\sigma_\epsilon$ for each sample defined in Sec~\ref{sec:galsamples}, as a function of $r$-band magnitude cut. 
We show $\sigma_\epsilon/\sqrt{n_\mathrm{gal}}$ for LS DR9 $R_h>1.\arcsec0$ (sample 1), BGS Any $R_h>1.\arcsec0$ (sample 3), and BGS Bright $R_h>1.\arcsec0$ (sample 5) in Fig.~\ref{fig:SN_over_ngal}, where we also annotate the LSST Y1, Y6, and Y10 levels. 
The other three samples with hlr $> 1.^{\prime\prime}5$ are less compelling than their hlr $>1.^{\prime\prime}0$ equivalent and thus are not shown here.
Overall, those three samples have similar $\sigma_\epsilon/\sqrt{n_\mathrm{gal}}$ as LSST Y1, albeit over very different redshift ranges. 
We note that although the shape noise degrades quickly as the sample becomes fainter, including all the faint galaxies in those samples is still beneficial, and we do not down-select the six samples in the following analyses. 
The final sample-averaged shape noises for samples 1--6 are around 0.071, 0.050, 0.047, 0.039, 0.041, and 0.037 respectively, based on our simulated KL measurements. These numbers serve as reasonable estimates of shape noise that account for realistic data quality and sample properties.

\section{Simulated Likelihood Analyses of very-low-redshift $S_8$ constraints}
\label{sec:results}

\subsection{Science case motivation}
In this paper we explore the science case of a high-precision ``very-low-redshift $S_8$'' measurement, which is motivated by the emerging $S_8$ tension between the primary CMB observables and low redshift clustering and weak lensing measurements. Recent results comparing Planck data with data from DES, KiDS, HSC, or BOSS, indicate a 2--3$\sigma$ difference between low and high redshift constraints on $S_8$~\citep{P18A6,KiDS1000_3x2pt,TKR+21,KNTM22,DES_Y3_3x2pt,HSC_Y3_SMM+23,LHL+23},. This tension is slightly reduced for the case when comparing Planck to CMB lensing measurements, e.g., from Planck itself or ACT, or when including CMB lensing information in the low-redshift datasets of weak lensing and galaxy clustering. \citep{DES_Y3_6x2pt_III_23,CML22,XEM+23,ACT_DR6_CMBL,FKM+23}

If this emerging picture is confirmed by future surveys (in particular by DESI, LSST Y1, ACT, SPT), it is not unlikely that the source of the $S_8$ tension is located at very low redshifts. This very low-redshift regime at $z<0.15$ is an excellent target for a KL initiative since constraints from galaxy clustering are limited by lack of volume and traditional weak lensing/cosmic shear analyses are integrated measurements that are mostly sensitive to significantly higher redshifts. While KL is an integrated measurement as well, its low shape noise per galaxy can efficiently probe the low redshift range without including higher redshift galaxies. 

We therefore consider a tomographic weak lensing angular power spectrum analysis of the 14,000 deg$^2$ DESI-KL survey. We break the KL samples into four tomographic bins with equal number density per tomographic bin. As a comparison, we also consider two LSST weak lensing scenarios as defined in~\citep{DESC_SRD}: 12,300 deg$^2$ LSST Year 1 with 11.1 gal/arcmin$^2$ and 14,300 deg$^2$ LSST Year 10 with 27.7 gal/arcmin$^2$, both with ten equal-density tomographic bins and a shape noise of 0.37~\citep{Chang13}. 

\subsection{Methodology}

We use \CL~\citep{EKS+14,KE17} for computing the data vector $\bm{D}$ and its covariance $\mat{C}$ at a fiducial point in parameter space, following the equations for cosmic shear power spectra and covariances described in Sec. \ref{sec:KLbasic}. The $C^{ij}_{\kappa\kappa}(\ell)$ is evaluated at 15 logarithmic bins in $20\leq\ell\leq3000$.
We split the $\sigma_8$ parameter into two independent parameters that are sensitive to clustering above and below redshift of 0.15, denoted as $\sigma_8(z\ge0.15)$ and $\sigma_8^{\mathrm{low}-z}$ respectively in the following analyses. We only consider these two cosmological parameters in our analysis and we will comment on this fact further below in Sec. \ref{ssec:results}.

Regarding the systematics, we include shear calibration bias, redshift uncertainties, and baryonic feedback in both DESI-KL and LSST-WL analyses. Since KL measures cosmic shear and intrinsic shape separately for each galaxy, it is immune to intrinsic alignment (IA) theoretically. Thus we assume perfect IA control for the KL sample while mitigating the IA biases in LSST analyses. We refer to Appendix~\ref{sec:append_sys} for detailed settings of systematics modeling. 

The posterior probability of the parameter space (2D cosmological parameters $\bm p_\mathrm{co}=(\sigma_8^{\mathrm{low}-z},\, \sigma_8 (z \ge 0.15))$ plus nuisance parameters) is sampled by MCMC chains using Bayes' theorem assuming that a multi-variate Gaussian likelihood describes the statistical distribution of our data points
\begin{equation}
\label{eq:like}
\mathcal L (\bm D| \bm p) = N\times\exp \biggl( -\frac{1}{2} \underbrace{\left[ (\bm D -\bm M)^\mathrm{T}\cdot \mat{C}^{-1}\cdot(\bm D-\bm M) \right]}_{\chi^2(\bm p)}  \biggr) \,,
\end{equation}
where the model vector $\bm M$ is a function of cosmology and nuisance parameters, and the normalization constant $N$ can be ignored under the assumption that the covariance is constant in the parameter space~\citep[see][for discussions]{ESH09,KAF19,AHE24}.

\begin{figure*}
    \includegraphics[width=0.48\linewidth]{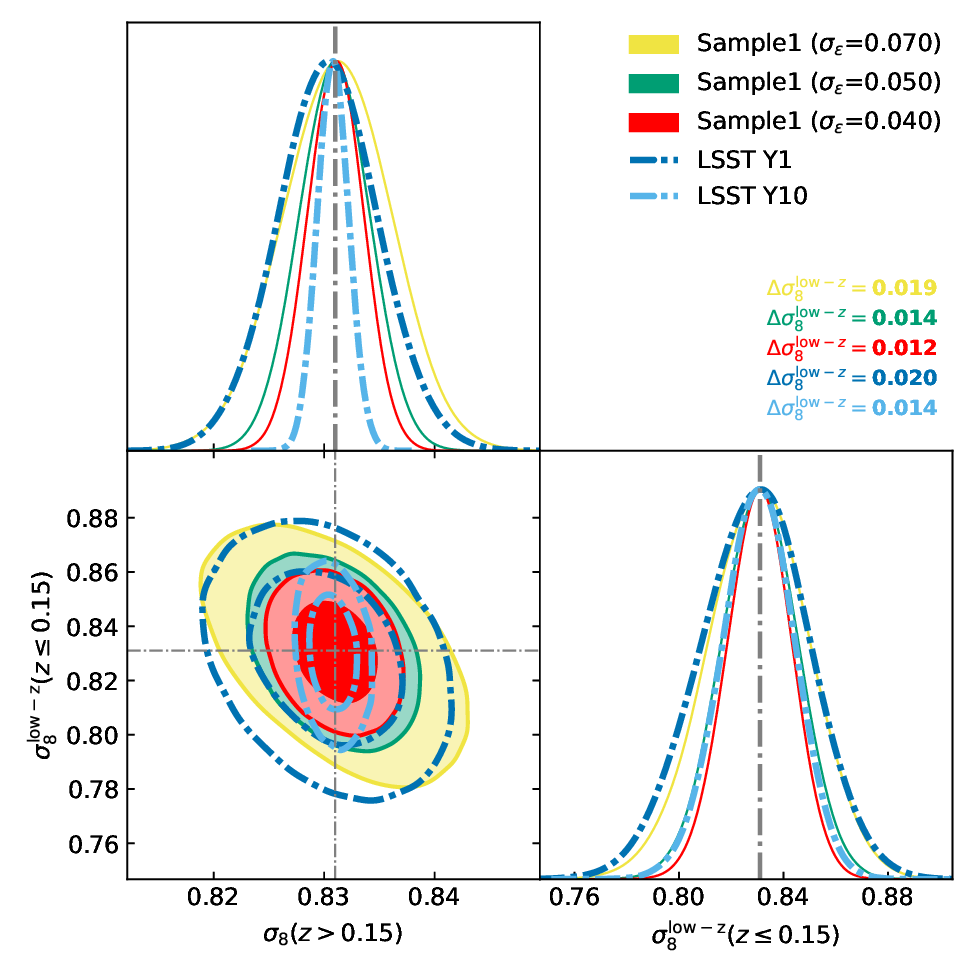}
    \includegraphics[width=0.48\linewidth]{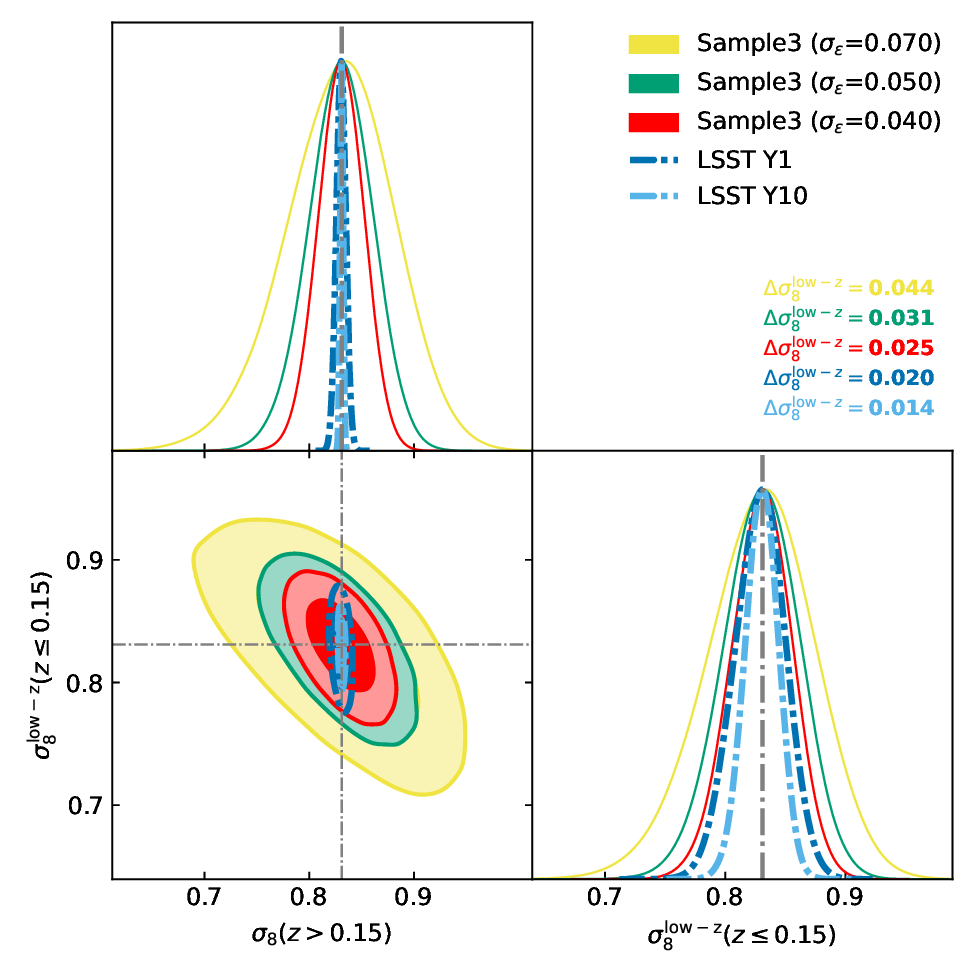}
    \caption{Constraints on $\sigma_8$ above and below $z=0.15$ for different KL-DESI samples, i.e. the extended BGS-like sample selected from LS DR9 (left) and BGS Any sample (right), and for different shape noise levels (red $\sigma_\epsilon=0.040$, green $\sigma_\epsilon=0.050$, yellow $\sigma_\epsilon=0.070$). We also show results from LSST Y1 and Y10 traditional cosmic shear (dark and light blue).}
    \label{fig:sigma8}
\end{figure*}

\subsection{Results}
\label{ssec:results}
We consider several DESI-KL surveys that all cover 14,000 deg$^2$ but have different galaxy samples (see Fig. \ref{fig:zdist}) and shape noise levels. The primary variations are sample size, redshift distribution, and shape noise level that can be achieved. These variables are directly connected to questions of survey strategy and any potential DESI-KL survey will have to optimize the trade space of exposure time, number of fibers per galaxy measurement, available galaxies in the field of view, and the quantities that directly affect constraining power. 

The results of our simulated analyses are depicted in Fig. \ref{fig:sigma8}. From left to right we show results for our samples 1 and 3, i.e the Legacy Survey sample with cuts in magnitude $r \leq 22$ and half-light radius hlr $\geq$ 1 arcsec, and for the BGS Any sample (hlr $\geq$ 1.0 arcsec), which yield $n_{\rm{gal}} = 0.53/\rm{arcmin}^2$, and $n_{\rm{gal}} = 0.18/\rm{arcmin}^2$, respectively. The different colors in Fig. \ref{fig:sigma8}, correspond to different shape noise levels, specifically, $\sigma_\epsilon=0.04$ (red), $\sigma_\epsilon=0.05$ (green), $\sigma_\epsilon=0.07$ (yellow), which can be compared to the expected shape noise levels for each galaxy sample in Fig. \ref{fig:SN_over_ngal} and Sec.~\ref{sec:average_sn}. 
In addition to the DESI-KL scenarios, we also show LSST Y1 (dark blue, dotted-dashed) and LSST Y10 (light blue, dotted-dashed) analyses using the redshift distributions and galaxy number densities as defined in \citep{DESC_SRD}. We note that the systematics quoted in~\citep{DESC_SRD} is a little bit optimistic and is challenging to achieve. 

The most constraining contour, the Legacy Survey sample (sample 1) with $\sigma_\epsilon=0.04$, gives tighter constraints on $\sigma_8^{\mathrm{low}-z}$ compared to even LSST Y10. Reducing the shape noise for this sample to $\sigma_\epsilon=0.05$ returns constraints that are located between LSST Y10 and Y1 constraining power, and going further to $\sigma_\epsilon=0.07$ is still competitive with LSST Y1. The results of our shape noise simulation in Fig. \ref{fig:SN_over_ngal} however demonstrate that a 600s exposure time is not sufficient to achieve $\sigma_\epsilon=0.04$ and that the more realistic constraining power of the Legacy Survey sample will be closer to the yellow contours. 

For the BGS Any sample (sample 3), the reduced number density of galaxies significantly degrades constraining power compared to the full Legacy Survey sample. 
Nevertheless, the cases of $\sigma_\epsilon=0.04$ and $\sigma_\epsilon=0.05$ are not significantly degraded compared to LSST-Y1. 
The shape noise simulation results in Fig. \ref{fig:SN_over_ngal} indicate that a KL measurement with the BGS sample can achieve constraining power located between the red and green, even with only 600s exposure time. 
We consider this BGS Any scenario to be an interesting concept to explore further, given the limited amount of time required to implement such a survey (see Sec.~\ref{ssec:implementation}).

The fact that we simulate a 2D $\bm p_\mathrm{co}=(\sigma_8^{\mathrm{low}-z}, \sigma_8 (z \ge 0.15))$, while keeping all other cosmological parameters fixed impacts DESI-KL and LSST results very differently. 
In a more realistic analysis that opens up further cosmological parameter dimensions, a DESI-KL survey will struggle to constrain several cosmological parameters, in particular, $n_s, h_0, \Omega_b$ due to lack of volume considered, and its number density is also low as a weak lensing survey, while LSST will do much better in self-constraining these parameters.
However, an immediate boost to the performance of a DESI-KL survey is to include the 3D clustering of the same galaxy sample since the spectra information is a prerequisite, and the fact that the lensing and clustering are measured from the same sample further shrinks the uncertainty due to their shared systematics~\citep{SEB24}. Another upgrade to a DESI-KL survey is replacing the imaging data with the UNIONS survey, which is much deeper ($r$-band depth 24.9) than the Legacy Survey and also covers the northern DESI footprint. As we mentioned in Sec.~\ref{ssec:synergies}, an $r$-band 24.9 depth imaging quality can decrease the shape noise by $\sim 20$ percent. In that case, the shape noise of the BGS Any sample is around 0.38 and its performance is the same as LSST Y1. 

Also, we note that a DESI-KL survey is very robust against the common weak lensing systematics. In an analysis that fixes all the nuisance parameters, the derived constraining power on $\sigma_8^{\mathrm{low-}z}$ is the same as the baseline analysis, while the LSST Y1 and Y10 analyses degrade a lot by opening up systematics dimension, and could degrade more if the systematics adopted in this work can not be met. 

Consequently, the tomographic weak lensing power spectrum is not an optimal way to extract the information encoded in a DESI-KL survey. A more meaningful comparison will have to include multi-probe analyses and study the trade space of systematics and external probes, which is beyond the scope of this paper and we defer to future work.

\subsection{DESI-KL survey implementation}
\label{ssec:implementation}
A 14,000 deg$^2$ DESI-KL survey based on the largest galaxy sample considered in this paper (see Fig.~\ref{fig:zdist}), i.e. the Legacy Survey sample ($r \leq 22,\,\mathrm{hlr}\geq 1^{\prime\prime}$, $n_\mathrm{gal}=1896$ galaxies deg$^{-2}$) would require 3532 hours of DESI dark time. These calculations assume four offset fibers across the galaxy with 600 seconds nominal exposure time. 
Given the amount of dark time of DESI per year, i.e. 1383 hours~\citep{SKS+23}, and accounting for an average dust and air mass extinction of 1.51, such a survey would consume all of the DESI dark time over 3.85 years. Variations that only assume two fibers per galaxy, a smaller survey footprint, or an increased/decreased exposure time are straightforward to calculate from this estimate. 

If one were to target the BGS Any sample ($\mathrm{hlr} \geq 1^{\prime\prime}$, 649 galaxies deg$^{-2}$) the required survey time reduces to 1202 hours of DESI dark time, which corresponds to $\sim$ 16 months. Several differences between these two scenarios should be noted:
\begin{itemize}
    \item While a fair fraction of the BGS Any galaxies can be targeted during bright time, the bulk of the fainter galaxies in the Legacy Survey sample will require dark time observations.
    \item Since all BGS Any galaxies already have a central fiber measurement this allows us to identify galaxies with bright emission lines and to rank them as a function of emission line flux. This knowledge will allow us to optimize target selection for bright and dark times and hence overall survey strategy. For the majority of the Legacy Survey sample, however, this estimate will be solely based on galaxy colors and morphology. 
\end{itemize}

An actual implementation of a DESI-KL survey of course depends on several external considerations, in particular on the overall integration of the different science cases and their prioritization in a coherent survey strategy. We note that DESI-KL is an excellent candidate for a spare fiber program since the target density is extremely high. 

\subsection{Synergies with LSST}
\label{ssec:synergies}
We stress that cosmological constraints from LSST weak lensing and the DESI-KL survey idea are highly complementary. Given that DESI-II will primarily cover the northern sky and LSST will survey the south, a DESI-KL initiative allows for an interesting comparison and ultimately combination of the two constraints. 

\begin{figure}
    \centering
    \includegraphics[width=\linewidth]{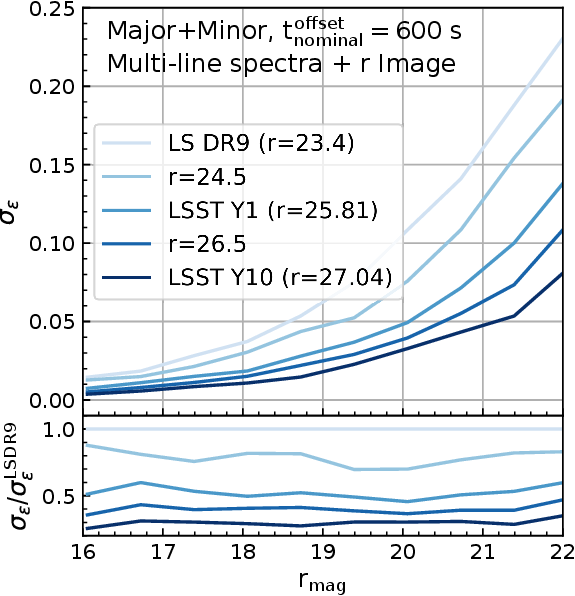}
    \caption{Shape noise as a function of deeper photometric coverage. Here we show how $\sigma_\epsilon$-$r$ relation is impacted by the survey depth of the imaging component. Lines from light to deep blue assume $r$-band survey depth of 23.4 (Legacy Survey DR9 level), 24.5, 25.81 (LSST Y1 level), 26.5, and 27.04 (LSST Y10 level). This indicates the gains in shape noise that can be obtained in the overlapping area of DESI and LSST and of course also if DESI overlaps with other photometric surveys of relevant depth. Here we assume four offset fibers are placed along major and minor axes of the galaxy with a nominal exposure time of 600 seconds. The shape noise is derived by fitting $r$-band photometry image and spectra of $\mathrm{H}\,\alpha$, $[\mathrm{O}\,\textsc{ii}]$, $[\mathrm{O}\,\textsc{iii}]4960$, and $[\mathrm{O}\,\textsc{iii}]5008$.}
    \label{fig:SN_photodepth}
\end{figure}

In the overlap region of LSST and DESI-II the benefits are even more tangible. On the LSST side, spectroscopic information will allow for the well-known benefits of better photo-$z$ calibration and improved control of intrinsic galaxy alignment uncertainties. On the DESI-KL side, we simulate the impact of improved (deeper) photometric imaging on the KL measurement technique and summarize our results in Fig. \ref{fig:SN_photodepth}. We find that access to LSST Y1 photometric information ($r\approx25.81$ mag) can further reduce the residual KL shape noise by approximately a factor of 2. This increases to a factor of 3 improvement when using LSST Y10 photometric information. 
Of course, instead of LSST, any other deep imaging surveys that cover the northern sky (e.g. Euclid and UNIONS) can be used to boost the KL shape noise gains. 

\section{Conclusions}
\label{sec:conc}
The $\sim$2 $\sigma$ difference in $S_8$ between primary CMB measurements from Planck and low redshift observations from weak lensing and galaxy clustering is one of the primary hints of tensions in the $\Lambda$CDM model. This difference generally becomes less pronounced when CMB lensing (from Planck and ACT) is folded into the analysis, either in combination with galaxy lensing and clustering, or when comparing to Planck primary CMB measurements directly. This motivates the possibility that the origin of the $S_8$-tension may be located at very low redshifts. 

In this paper, we explore the prospects of a Kinematic Lensing survey with the Dark Energy Spectroscopic Instrument (DESI-KL) that would target this low redshift regime and attempt a high-precision $S_8$ measurement. 
DESI-KL is based on the idea of breaking the degeneracy between intrinsic shape and cosmic shear with both photometry and kinematic information of disk galaxies. 
The kinematic information of a disk galaxy is measured by pointing additional fibers along the major and minor axes of the galaxy.

Our simulation program to assess the constraining power of such a survey is based on two main pillars: 1) Firstly, we simulate the expected shear measurement uncertainty of a realistic DESI-KL mock observation for a variety of different galaxy samples, and therefore the resulting shape noise. 2) Secondly, we run simulated likelihood analyses that forecast the constraining power on a low-$z$/high-$z$ $\sigma_8$ split parameter set, where the low-$z$ corresponds to $z \leq 0.15$ and high-$z$ to $z > 0.15$. 

We focus on exploring an extended Bright Galaxy Survey sample selected from the DESI Legacy Survey (sample 1, $n_{\rm{gal}}=0.527\,\mathrm{arcmin}^{-2}$) and the Bright Galaxy Survey sample (BGS Any or sample 3, $n_{\rm{gal}}=0.180\,\mathrm{arcmin}^{-2}$), where the latter already have spectra at the center of the galaxy. 
The relevant quantity for statistical noise in cosmic shear surveys is $\sigma_\epsilon/\sqrt{n_{\rm{gal}}}$, which motivates the idea to include larger galaxy samples in the analysis. Our KL inference pipeline however shows that the residual shape noise in the KL measurement strongly depends on the magnitude of the galaxy or signal-to-noise. 
Consequently, we find that statistical gains, e.g. when going from the BGS sample to sample 1, are less significant and less economically efficient given the threefold longer total exposure time required.

We compare the $\sigma_\epsilon/\sqrt{n_{\rm{gal}}}$ values of the extended BGS sample and the BGS Any sample to corresponding numbers that traditional weak lensing from LSST Y1 (12,300 deg$^2$, $n_{\rm{gal}}=11.1\, \mathrm{arcmin}^{-2}$) and LSST Y10 (14,300 deg$^2$, $n_{\rm{gal}}=27.7\,\mathrm{arcmin}^{-2}$) will achieve, albeit with a galaxy sample spread out over a much larger range in redshift. 
While the statistical power of LSST Y10 remains unmatched, both samples cross below the LSST Y1 statistical noise level. Generally, we find that our KL measurement precision does not increase rapidly with exposure time or with the number of fibers placed on the galaxy (we consider two and four-fiber spectra in our analysis), but these statements need to be revisited when analyzing a pilot dataset. Given these findings, we suggest that a DESI-KL survey focused on the BGS Any sample can provide a formidable, localized measurement of structure formation at low redshift. 

We confirm this idea by running simulated likelihood analyses that sample the two-dimensional cosmological parameter space of $\bm p_\mathrm{co}=(\sigma_8^{\mathrm{low}-z}, \sigma_8 (z \ge 0.15))$ along with systematics. 
We find that a BGS Any KL survey can indeed constrain low redshift $\sigma_8$ at a similar level to LSST Y1, especially when deeper northern sky imaging is available (e.g. UNIONS). 
To reach LSST Y10 constraining power, a KL survey focused on the extended BGS sample with significantly longer exposure time than 900s may be required. 

Overall we find that a DESI-KL survey that targets low-$z$ structure formation with the BGS Any galaxy sample is an interesting option that should be explored further. Such a survey will be quasi-immune towards the standard suite of systematics that haunt traditional weak lensing measurements (shear and photo-$z$ calibration, and intrinsic alignment). However, the small volume of such a survey will require additional information to constrain cosmological parameters that govern the geometry of our Universe ($n_s,\,h,\,\Omega_b$). Luckily, these parameters are tightly constrained by the DESI clustering and BAO programs, which enables a nice synergy of DESI internal observations. We stress that DESI-KL and LSST are highly synergistic endeavors. 
Together these can enable a lensing program that spans the entire northern and southern galactic sky, which are very different in terms of instrumentation and measurement technique, and that consequently enable a variety of studies that compare and ultimately combine both information sources. 

Kinematic Lensing with DESI can be a highly synergistic program with several other DESI science cases. The combination of low redshift lensing measurements and peculiar velocity information enables the design of formidable tests for modified gravity, similar to the combination of clustering and peculiar velocities that was mapped out in \cite{DESIPV23}. Furthermore, any dark matter studies that use lensing to measure the mass profile of dwarf galaxies \cite{LSL+20,LLG+24,TAW+23} will greatly benefit from the enhanced signal-to-noise. If successful, such a DESI-KL program can enable such measurements on individual dwarf galaxies rather than stacks.

The findings in this paper need to be validated via a pilot study on actual DESI (peculiar velocity) data that informs the exact trade space of exposure time, fiber position(s), and galaxy properties, which we defer to future work.

\begin{acknowledgments}
The authors thank David Schlegel, Martin White, and Kyle Dawson for valuable discussions on this work. JX, TE, EW are supported by the Department of Energy grant DE-SC0020215 and by funding from UArizona Research, Innovation \& Impact (RII). The simulations in this paper use High Performance Computing (HPC) resources supported by the University of Arizona TRIF, UITS, and RDI and maintained by the UA Research Technologies department. 
\end{acknowledgments}

\appendix

\section{Exact Selection Criteria}
\label{sec:append_ts}
Here we show the exact target selection criteria used in Sec.~\ref{sec:galsamples}

\paragraph*{Criterion 1.a} Similar to the BGS sample target selection~\citep{BGS_selection2}, we first impose the star-galaxy separation
\begin{equation}
    (\texttt{REF\_CAT!=G2})\,|\,(G_\mathrm{Gaia}-r_\mathrm{raw}>0.6),
\end{equation}
then we cut on fiber magnitude
\begin{equation}
r_\mathrm{fiber}<\begin{cases}
    22.9+(r-17.8) & \text{for } r<17.8,\\
    22.9 & \text{for } 17.8<r<22,
\end{cases}
\end{equation}
and require it to be observed in $grz$ bands
\begin{equation}
    (\texttt{NOBS\_G}>0)\, \&\, (\texttt{NOBS\_R}>0)\, \&\, (\texttt{NOBS\_Z}>0).
\end{equation}
We also discard objects with extreme color by requiring
\begin{equation}
\begin{aligned}
    -1 &< g-r &< 4, \\
    -1 &< r-z &< 4,
\end{aligned}
\end{equation}
and cut out bright objects that satisfy
\begin{equation}
    (r>12) \,\&\, (r_\mathrm{fibertot}<15).
\end{equation}

\paragraph*{Criterion 1.b} This criteria includes galaxies in BGS Bright and BGS Faint. Since BGS science validation spectrum are available when we plan the KL survey, we also require the sample to have good redshift measurement
\begin{equation}
\label{eqn:ts_good_z}
\begin{aligned}
    (\texttt{ZWARN}==0) \,&\&\, (\texttt{SPECTYPE}==\texttt{GALAXY}) \,\&\,\\ (\Delta\chi^2>40) \,&\&\, (\texttt{ZERR}<0.0005\,(1+\texttt{Z})).
\end{aligned}
\end{equation}
\paragraph*{Criterion 1.c} Same as 1.b except here we require the galaxy is in BGS Bright.

\paragraph*{Criterion 2.a} Here we require that the morphology is classified as late-type galaxies
\begin{equation}
\texttt{MORPHTYPE}=\begin{cases}
    \texttt{EXP} \\
    \texttt{REX} \\
    \texttt{SER }\text{for } \texttt{SERSIC}\leq 2
\end{cases},
\end{equation}
and the half-light radius is larger than 1 arcsec
\begin{equation}
    \texttt{SHAPE\_R}\geq 1.^{\prime\prime}0.
\end{equation}

\paragraph*{Criterion 2.b} Same as 2.a except we require the half-light radius to be $\texttt{SHAPE\_R}\geq 1.^{\prime\prime}5$.

\section{Systematics Modeling}
\label{sec:append_sys}

\paragraph*{Shear calibration bias} We consider multiplicative shear calibration bias $m^i$ in this work
\begin{equation}
    C^{ij}_{\kappa\kappa}(\ell)\rightarrow (1+m^i)(1+m^j)C^{ij}_{\kappa\kappa}(\ell),
\end{equation}
and we impose Gaussian priors centered at zero for $m^i$. We set the standard deviation of $m^i$ to 0.013 (LSST Y1) and 0.003 (LSST Y10) based on~\citep{DESC_SRD}. 
We note that the KL sample is much brighter and larger than the traditional weak lensing sample, and we assume a standard deviation of 0.0004 for it.

\paragraph*{Redshift uncertainty} We model the redshift uncertainty as a Gaussian smoothing with mean of $\Delta_{z}^{i}$ and width of $\sigma_{z}^{i}(1+z)$~\citep[e.g., see][]{XEH+23}. We assume $(\Delta_{z}^{i},\,\sigma_{z}^{i})=(0,\,0.05)$ for LSST Y1 and Y10, and we impose Gaussian priors on $\Delta_{z}^{i}$ and $\sigma_{z}^{i}$. For LSST Y1, $\Delta_{z}^{i}\sim \mathcal{N}(0,0.002)$ and $\sigma_{z}^{i}\sim \mathcal{N}(0.05, 0.006)$. For LSST Y10, $\Delta_{z}^{i}\sim \mathcal{N}(0,0.001)$ and $\sigma_{z}^{i}\sim \mathcal{N}(0.05, 0.003)$~\citep{DESC_SRD}. For DESI-KL, note that the sample has spectroscopic redshift, we set $(\Delta_{z}^{i},\,\sigma_{z}^{i})=(0,\,0.0005)$, which corresponds to the target selection criteria in Eq.~(\ref{eqn:ts_good_z}). We set their priors to $\Delta_{z}^{i}\sim \mathcal{N}(0,0.0001)$ and $\sigma_{z}^{i}\sim \mathcal{N}(0.0005, 0.0001)$.

\paragraph*{Baryonic feedback} We use principal component analysis (PCA) method to mitigate the baryonic feedback bias in weak lensing power spectrum~\citep{EKD+15,HEM+19,HEM+21,XEM+23}. 
The principal components (PCs) are reduced from a suite of different hydrodynamical simulations: Horizon-AGN~\citep{dpp16}, Illustris/IllustrisTNG~\citep{wsp18,psn18}, Eagle simulation~\citep{scb15}, Massiveblack-II~\citep{kdc15} and the OWLS AGN simulation~\citep{sdb10,dsb11}. We marginalize over the first 2 PCs with Gaussian priors on $Q_1\sim\mathcal{N}(0,40)$ and $Q_2\sim\mathcal{N}(0,10)$ for both LSST-WL and DESI-KL. 

\paragraph*{Intrinsic alignment}
We use nonlinear alignment~\citep[NLA][]{HS04,BK07,KEB16} model to mitigate the IA bias in LSST-WL analyses. Four parameters are introduced to control the NLA model: the IA amplitude $A_{\mathrm{IA}}\sim \mathcal{N}(5.92, 3.0)$, the luminosity dependency $\beta_{\mrm{IA}}\sim \mathcal{N}(1.1, 1.2)$, and the redshift dependency $\eta_{\mrm{IA}}\sim \mathcal{N}(-0.47, 3.8)$ for $z\leq0.3$ and $\eta_{\mrm{IA}}^{\mrm{high}\textit{-}z}\sim \mathcal{N}(0, 2.0)$ for $z>0.3$~\citep{JMA+11}.
We do not include IA uncertainties in KL analyses but see~\citep{HKX+24} for a discussion of IA-like systematics that only affects KL, which is likely to be negligible.

\bibliography{main}

\end{document}